# Web of Science: showing a bug today that can mislead scientific research output's prediction

Pablo Diniz Batista[1]*,

Igor Marques-Carneiro[1],

Leduc Hermeto de Almeida Fauth[1] and

Márcia de Oliveira Reis Brandão[1]

[1]Brazilian Center for Research in Physics
Rua Dr. Xavier Sigaud, 150 - Urca - Rio de Janeiro - RJ - Brasil - CEP:22290-180

*Corresponding author: P. D. Batista

55 21 21417589

## Acknowledgments

The authors acknowledge Coordination for Improvement of Graduate People (CAPES) that allows the access to Web of Science database to Brazilian universities and research institutions. Data included here are derived from the Science Citation Index Expanded, prepared by Thomson Reuters, Philadelphia, Pennsylvania, USA, Copyright Thomson Reuters, 2016.

**Abstract**

As it happened in all domains of human activities, economic issues and the increase of people working in scientific research have altered the way scientific production is evaluated so as the objectives of performing the evaluation. Introduced in 2005 by J. E. Hirsch as an indicator able to measure individual scientific output not only in terms of quantity, but also in terms of quality, *h* index has spread throughout the world. In 2007, Hirsch proposed its adoption also as the best to predict future scientific achievement and, consequently, a useful guide for investments in research and for institutions when hiring members for their scientific staff. Since then, several authors have also been using the Thomson ISI Web of Science database to develop their proposals for evaluating research output. Here, using a software we have developed, we analyse more than 100 thousand articles and show that a subtle flaw in Web of Science can inflate the results of info collected, therefore compromising the exactness and, consequently, the effectiveness of Hirsch's proposal and its variations.

**Keyword:** Scientometrics, *h index*, Web of Science, predictability

## Introduction: A decade of *h* index

In 2005 Jorge Hirsch proposes a quantitative evaluation method that could be used to all research fields considering both quantity and quality (Hirsch 2005; Hirsch 2007). Since then we started discussing this indicator aiming to show that it is not merely a new mathematical value. It is important to notice that it was proposed in the United States and this is an aspect to be taken into account considering the quest for investing resources in areas that can give good economic results. According to the methodology proposed by Hirsch, to get to the *h* index, the first step is to organize all the articles published by a scientist in a sequence following a decreasing order of citation received by each article. Then, all the articles will be accounted, but some will be disregarded. That is, when performing the analysis, we have to decide when stop counting.

According to Hirsch, we should stop when the number of articles published by a scientist is smaller than the number of citations within this sequence. This methodology has a great deal of complexity but, as time goes by, researchers would internalize it and it would be possible to attribute a number to all scientists*:* the *h* index.

In order to analyse carefully Hirsch evaluation proposal, aiming to notice its nuances it is useful to start by the title of his first article: "An index to quantify an individual's scientific research output". His proposal is an attempt to deal with the issue of how to evaluate scientific research output, but it is also an opportunity to follow the birth of a measure within Mathematics. Another point to take into account is that researchers from many different fields of knowledge have been devoting themselves to the discussion of *h* index, showing that Scientometrics has been playing an important role for those directly, or not, involved in scientific research.

Hirsch proposal has spread throughout the world and nowadays a database calculates the *h* index for researchers that publish the results of their work in scientific journals indexed in the Science

Citation Index (SCI). The career of a scientist, taking into account the articles he published and the citations received, can be represented by a hyperbolic function.

Here the *h* index is determined when we draw a straight line from the origin to the curve of research output, so as the intersection between the line and the hyperbola is *the gold number*.

However, we could ask the following question: "how many lines – and consequently intersection points - could be drawn in this graphic?" We could say that there are many possibilities, so it is necessary to make a decision on which intersection point we will take for the analysis in a range of possibilities, because there is not yet a criterion to justify which one is the best line for a quantitative evaluation. Therefore, we could consider this as a first example that this approach is not completely neutral, at least when we try to create indicators to measure scientific productivity.

What is surprising is that the claim form neutrality is, at first, exactly what justify the wide acceptance and, consequently, adoption of Hirsch's proposal. According to it, the *h* index is able to join in a single indicator quality and quantity, allowing it to be used to measure research output of any research area (Hirsch 2005; Hirsch 2007).

Besides, in his second article, Hirsch starts, already in the title, with the question: "Does the *h* index have a predictive power?" That is, he extends his initial proposal of evaluation a researcher career from his past results to the possibility of predicting his future achievement based on the *h* index (Hirsch 2007). Furthermore, and very important, in his article Hirsch proposes that the indicator can be a very useful tool to guide the choice of scientific institutions when hiring someone to their staff. *h* index would allow to create a hierarchy among a group of candidates indicating which of them would give more results throughout the years.

## *h* index: an idea that has spread throughout the world

We started our work towards this indicator almost immediately after Hirsch article was published (Batista et al. 2006). In the present work we focus on two main goals: i) draw attention to the use by Thomson ISI of data collected from papers published since 1945; ii) discuss the proposals made by some authors to use bibliometric indicators as a tool to predict the performance of scientists. In order to achieve these goals, we show a counterexample in attempt to demonstrate that these indicators are still far from the ideal of neutrality of many scientists.

First, we show, in Figure 1-a**,** an overview of how the proposal presented by Hirsch has spread throughout the scientific world to an exciting speed. Since 2005, when it appeared for the first time, the article proposing the *h* index was cited at an increasing rate over the years, reaching almost 300 citations per year in 2013. Thus, the article has already garnered almost 2500 citations.

This shows the interest of our time in quantitative methods aiming to assess scientific output. The Figure 1-b shows the number of articles published up to now citing the first work of Hirsch focusing the first fifteen countries that contributed to the spread of *h* index throughout the scientific community. In addition, Figure 1-c shows that these citations came from different parts of the world as we can see in the geographical distribution map.

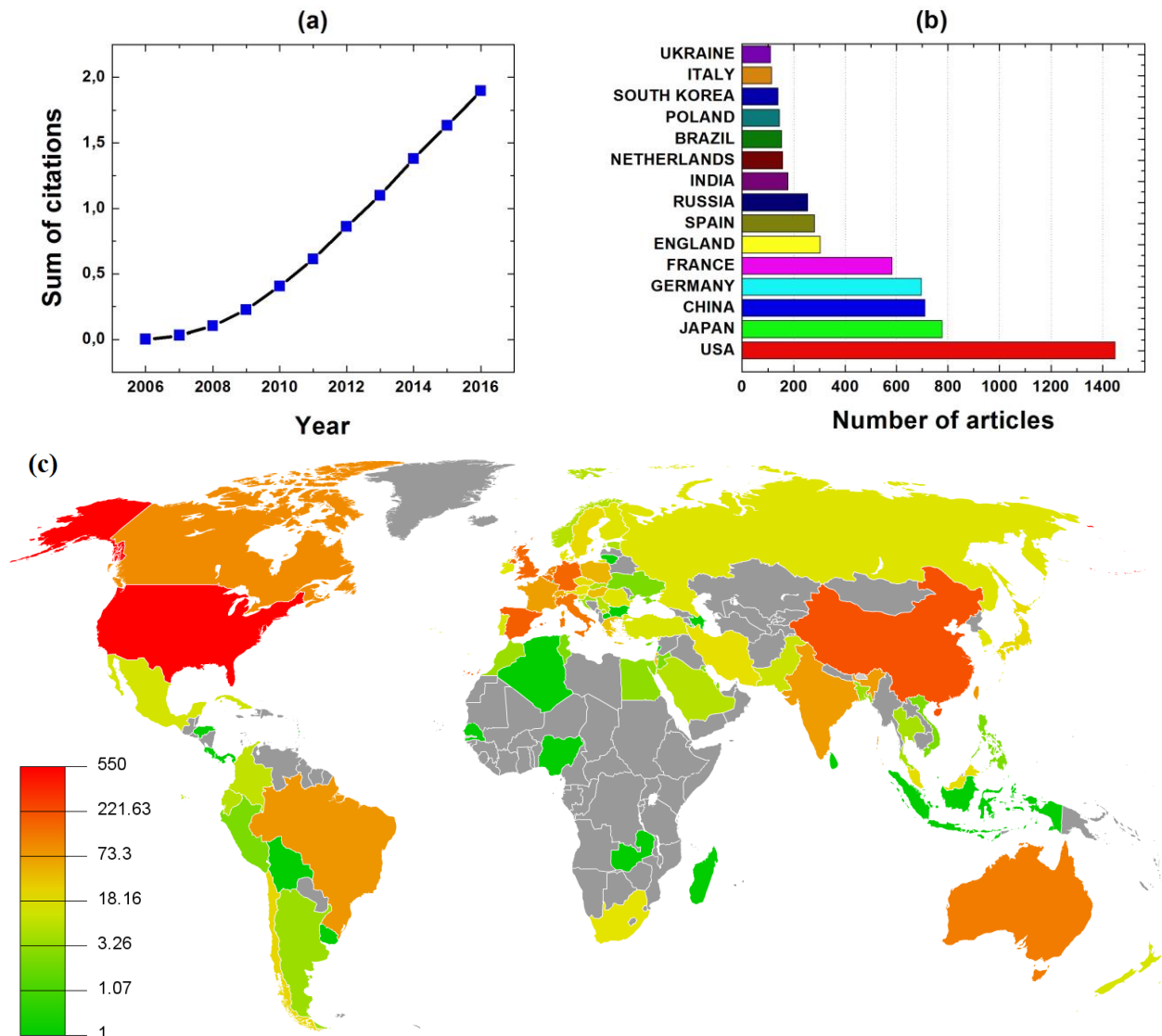

**Fig. 1** A decade of the *h* index in the world. (a) Number of citations that the first article published by Hirsch received since 2005. (b) Number of articles citing the first work of Hirsch focusing the first fifteen countries. (c) Geographical distribution of articles citing the first article of Hirsch on *h* index built from Arthur. GunnMap 2 - Global Chart Tool
http://lert.co.nz/map/

## *h* index : a temporal analysis

Eleven years after the publication of J. E. Hirsch's article proposing the *h* index as a bibliometric indicator capable of aggregating, in a single number, both the number of papers published and the quality of the research developed by a scientist, the present work performs a brief tour in the Web of Science intending to map the state of this indicator. We note that the proposed *h* index brought back a discussion on the possibility of finding indicators that may contribute to the assessment in science. There are different types of discussions on the Hirsch work making almost impossible for researchers who are not directly involved with the Scientometrics to keep track of all the issues that arose in the last ten years. On the other hand, it is important to follow the evaluation proposals since those indicators are being adopted not only as tool for evaluation but also as a guide in the decision-making process. Considering this, we would like to focus here on what Hirsch proposes in his second article (Hirsch 2007).

Hirsch first work on *h* index appeared in 2005. Although several articles have pointed out its weakness (Batista et al. 2006; Braun 2010), in short time Hirsch wrote three other articles trying to demonstrate its capacity to measure and evaluate research output. Among his articles, the second one is the best to allow us to identify the relationship between his scientometric proposal and a new order of performing science that began with Modernity.

In his second paper on *h* index, Hirsch proposes to use the indicator not only to classify scientists according to their past results, but also as a tool to predict their future scientific performance. Immediately after its publication, we note that several other articles were published addressing now the possibility of using this indicator as a representative measure able to predict future achievement of scientists considering only their scientific production, excluding, therefore, several factors probably not measurable (Daniel et al. 2012; Wang et al.2013).

Hirsch's proposal reminds us what has become famous as Laplace's Demon, especially when he asks in his second paper on *h* index:"which measure is better able to predict its [scientific achievement] future values?" (Hirsch 2007).

According to Laplace's philosophical propositions on probability, there would be "an intellect which at a certain moment would know all forces that set nature in motion, and all positions of all items of which nature is composed, if this intellect were also vast enough to submit these data to analysis, a single formula […], for such an intellect nothing would be uncertain and future, just like the past, would be present before its eyes" (Laplace 2014).

Aiming to analyse Hirsch's proposal, we observe that the bibliometric data collected from the Web of Science is not a reliable instrument for comparing scientists' performance because we detected a disregarded subtlety in the database. This flaw concerns the interpretation of how this indicator should be calculated (Hirsch 2005). In order to investigate how it can affect scientometric analysis, we have chosen to follow the career of the Nobel Prize in Physics 2010, Andre Geim (Geim 2011; Hancock 2001). This is not an arbitrary choice, since his example becomes interesting for the two major objectives of this work.

## The "bug" on *h* index from Web of Science

We know that, traditionally, the number of articles published and the total number of citations received by them is the basis for many scientific performance indicators that have been recently proposed. For this reason, the first option is to use the Web of Science to obtain the value of this indicator, within a time lag, in order to restrict the search according to a definite time interval of interest. The first step to analyse the evolution of bibliometric indicators is to search in the Web of Science considering the author name of a researcher. We decided to search for Geim's results from 1945 to 2016. The Web of Science presents the results allowing you to create a citation report as shown in figure 2-a.

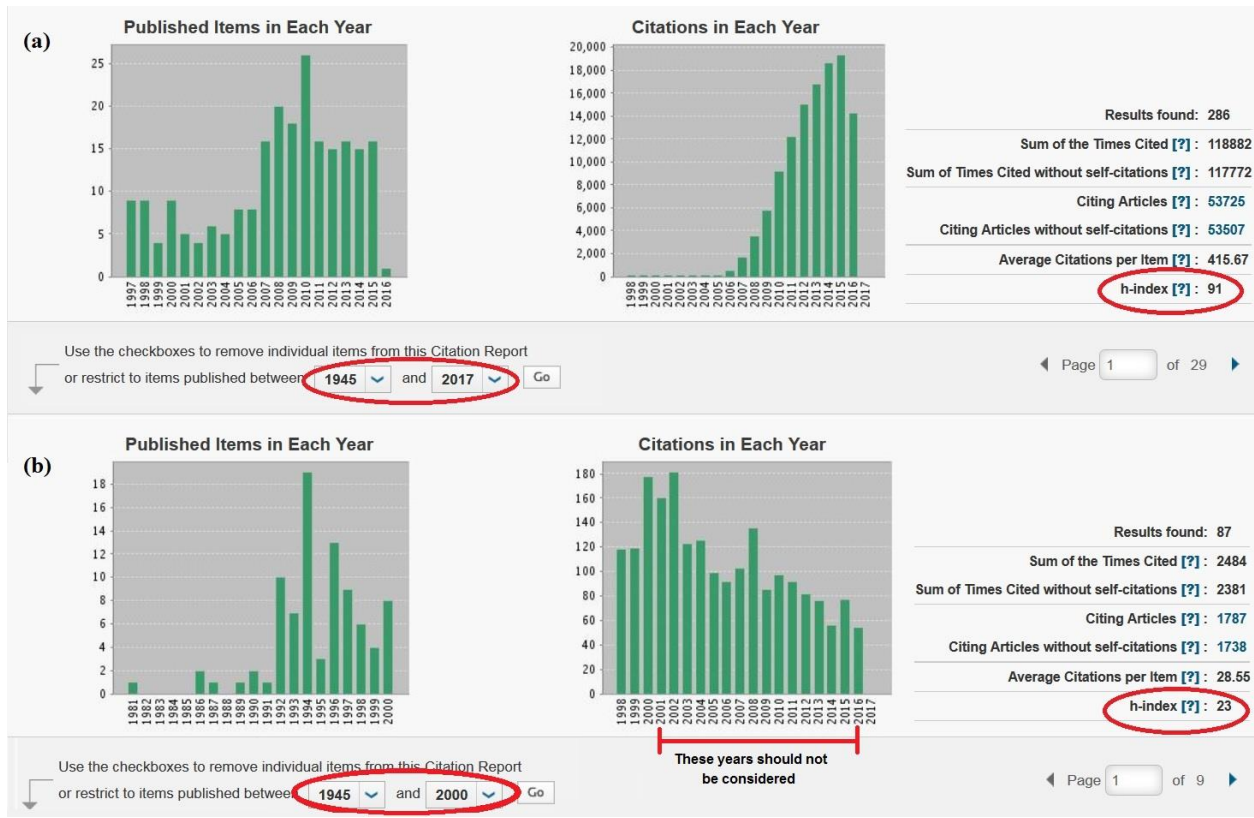

**Fig. 2** Temporal analysis of the *h* index using the Web of Science. In (a) we choose a time window between the years 1945 and 2016 obtaining a value of 91 for the *h* index. Then, to calculate the value of this indicator in 2000 we could simply change the time window as shown in (b), by selecting the years between 1945 and 2000. Note that the value of the *h* index is reduced to 23. We note, however, that the citations received are not restricted to calculate correctly the indicators, since all citations from 1945 to 2016 were accounted

The database presents six indicators to the user, including the *h* index. Just by changing the final year at the *TimeSpam* option, it would be easily possible to analyse the evolution of these indicators. For example, figure 2-b shows Geim's *h* index value in 2000. We first found that the tool correctly restricts the amount of articles published within the expected range (see *Published Items in Each Year* in Figure 2-b). However, it is observed that this does not happen to the citations garnered by the papers, once they are accounted until 2016 (see *Citations in Each Year* in Figure 2-b). That is, the results for all the indicators are inflated, since they also include the citations received between 2001 and 2016, therefore compromising the scientometric analysis.

We have developed a software able to get some piece of information automatically taking as input all citations received by the scientist we focus on here. In this case, we had to access data usually not available in the Web of Science in an attempt to put in question how this tool calculates *h* index over the years for each scientist. Note that the amount of information required to perform this operation is not available to the user in an accessible way in the Web of Science. For example, for Geim, in particular, over 100 thousand citations are accounted. To perform a temporal analysis of his career it would be necessary to fix the year of each one of these citations. We can do it, using Web of Science, accessing each one of the articles, what makes the task almost impossible when we investigate a scientist like Geim. The key point of our proposal here is that the software we developed automate the data extraction and processes it in a few minutes and made possible to us to analyse more than 100 thousand articles in the present work. In Figure 3 we show an overview of the software developed in python to access Web of Science data.

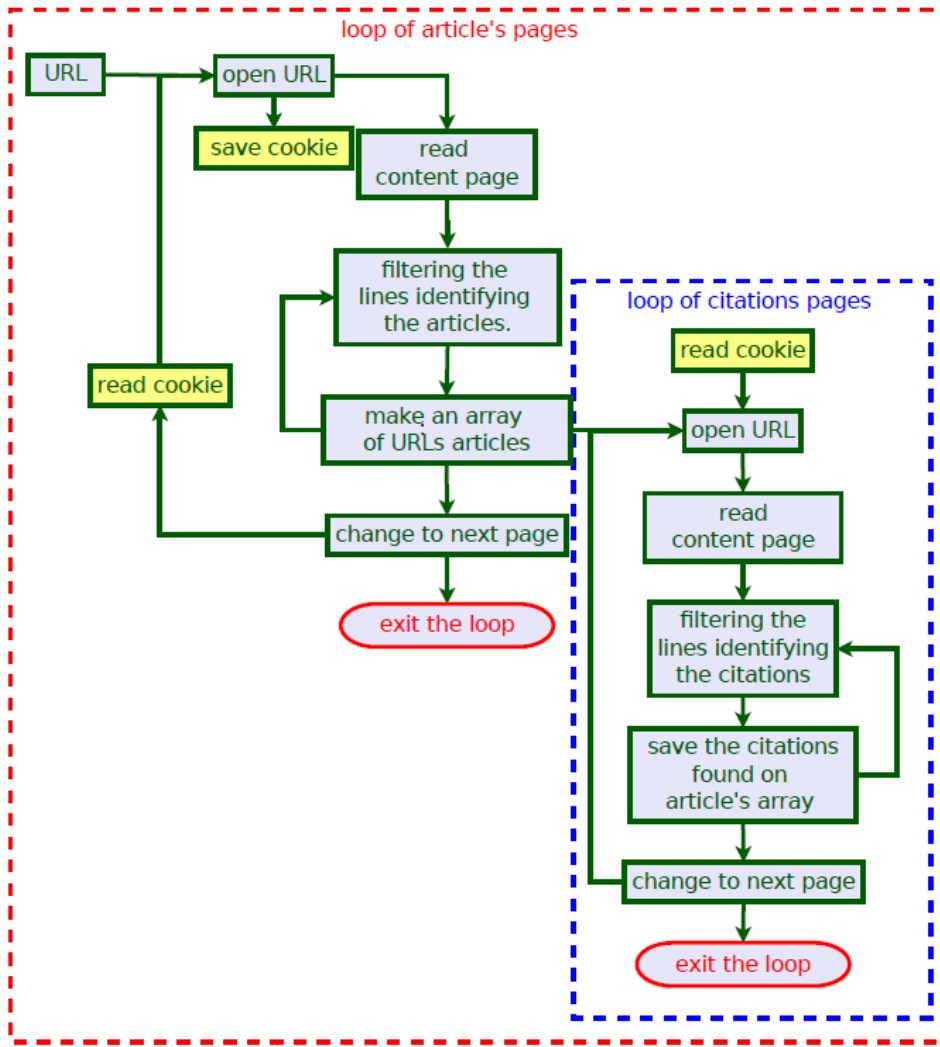

**Fig. 3** Flowchart of the software developed in python to access Web of Science data not available in order to quantify the results distortion since the tool does not properly restrict the number of citations according to the user-selected time window

The software is able to access all articles of a scientist in the Web of Science, collecting only the title, year of publication and all citations received by the article year by year. We discarded the rest of the information of the items and no piece of information was saved. It was created a text file for each article, containing the title on the first line, the year of publication of the article on the second line, the index identifier in the third row and then the *year of each citation one per line*. This software was developed in order to access the Web of Science site autonomously, that is, without human interference. To speed up the process, it runs on individual threads for each page of the list of items, reducing the final implementation time. The Web of Science site contains a security system that requires the cookie generated in the first access to access a second URL, this process is known as "session". To have access to this website, it was necessary to store the cookie information and reuse for each request.

Figure 4 shows the results for two scientometric analyses and the evolution of the values for this scientist throughout the years. It is easy to notice that the results show that the database increases in average 20% the value of his *h* index because it does not restrict correctly the number of citations received by each publication according to the user-selected period.

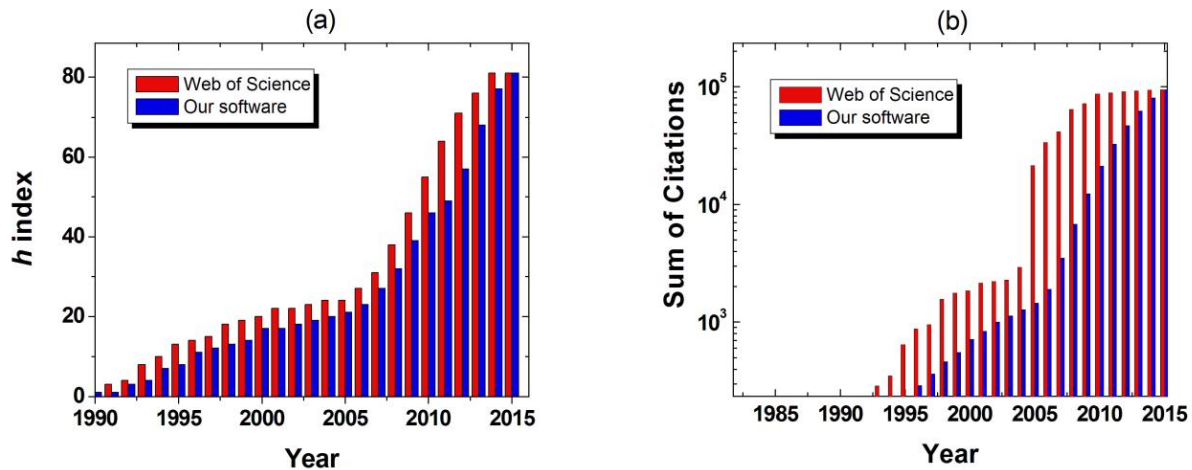

**Fig.4** Scientometric analysis of the career of the Nobel Prize in Physics in 2010: (a) *h*-index and (b) the total number of citations, both throughout time

## Andre Geim: a point outside the curve

In 2000 the physicist Andre Geim was awarded the Ig Nobel for his experiments with frog's levitation (Berry et al. 1997). It is important to consider that the Ig Nobel seems to be just a kind of joke towards scientific activity but it is able to provoke a profound reflection on many aspects involved in science. In fact, its motto is "first make people laugh, and then make them think". Moreover, ten years later, he receives the Nobel Prize in Physics for the isolation of graphene (Novoselov et al. 2004).

Until being awarded with the Ig Nobel, his papers had already a considerable number of citations. However, after the publication of the article on the isolation of graphene in 2004, the number of citations started to grow exponentially: it jumped from 9000 to 90000 within a very short period of time. The same also happened with his *h* index that went from 17 to 46 between 2000 and 2010.

Considering this, we would like to analyse the career of this scientist from the point of view of bibliometric indicators as shown in Figure 4. However, until the isolation of graphene in 2004 we can say that if we were to evaluate the performance of the researcher adopting the proposal made by Hirsch, we could not predict that this indicator would reach a value of 81 in 2015. We could make here an analogy with the fairytale The Prince Frog – based on a primary tale of Grimm Brothers – in which a princess kisses a frog and it turns into a prince: 'You have to kiss a lot of frogs before you find your handsome prince". Geim followed a certain random trajectory in his scientific career. His experiments with frog levitation described in (Berry et al. 1997) can be compared to the princess who kissed many frogs before finding her prince. In Geim's case, the article on graphene's isolation (Novoselov et al. 2004) was his prince frog.

It is remarkable that Andre Geim is the only scientist up to now to be awarded the Ig Nobel and the Nobel. In an interview after the Nobel, Geim talks about what he calls "Friday night experiments": "When you try something very elementary and try to go into one or another direction" (Geim 2011; Hancock 2001). No institutions sponsor these experiments, of course. That was what he was working on when he was nominated for the Ig Nobel.

Despite what it could represent being awarded it, Geim went on with his scientific career and ten years later, he got the Nobel for the isolation of graphene.

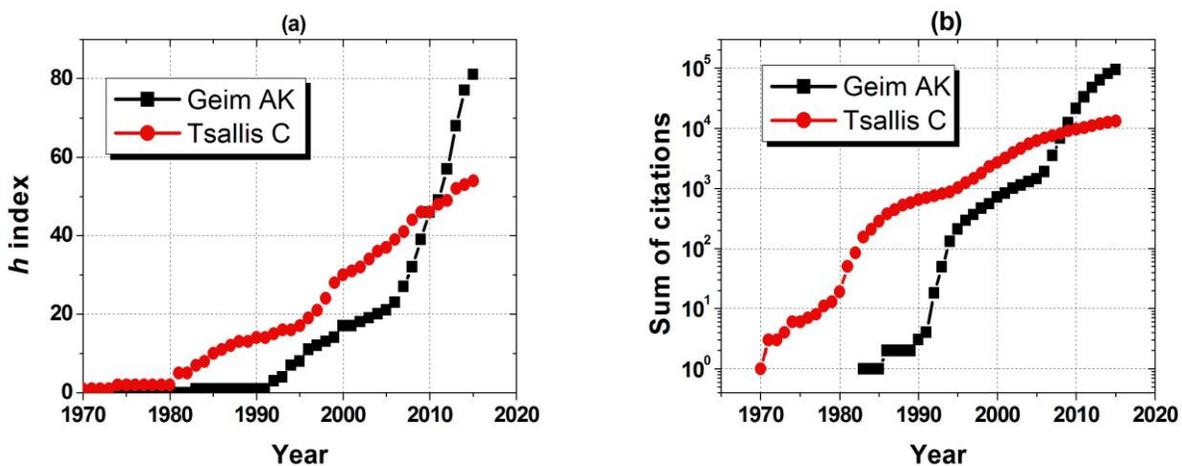

**Fig. 5** Comparative results for Geim and a Brazilian scientist. *h* index and the total number of citations throughout time for Geim AK and Tsallis C. It is possible to observe the turning point from 2004 in Geim's scientific career

Just to illustrate, Figure 5 shows the results of scientometric analysis when we compare Andre Geim to a Brazilian scientist (Tsallis 1988). We have chosen his example because he was at the fist position in a ranking we did in 2005. Taking as starting point Hirsh's proposal, the work proposed *h'*. This variation takes into account, when performing the analysis, the co-authorship in each article. For more details, see (Batista et al. 2006). It is possible to observe the turning point from 2004 on considering the results obtained for Geim's achievement for both indicators. Therefore, until 2004, according to our proposal here, it would not be possible to predict that Geim would surpass Tsallis in 2010 taking into consideration the results of those indicators presented in Figure 5. Nevertheless, as we know, after the publication of the articles on graphene, we note that the growth rate of those indicators for Andre Geim was significantly increased.

## Conclusions

We presented here a subtle flaw detected in 2015 in the Web of Science database during the analysis of the scientific career of the Nobel Prize in Physics in 2010 within the Scientometrics context (Zarka 2009). First, we note that the tool does not restrict appropriately the citations received by articles in accordance with the time lag set by the user during the search in the site when creating the *Citation Report*. Finally, we question the proposal recently made by some authors to use the *h* index as a statistical tool able to predict the performance of scientists. In order to achieve this goal, we have chosen as a methodological approach to present only an emblematic counterexample showing that this proposal would probably fail to predict that A. Geim would reach the results that allowed him to win the Nobel Prize. It would also probably fail to predict that his *h* index would reach an *h* index of 82 by 2015. We also demonstrated this through the example of a Brazilian researcher scientific output. This way, we intend to point out the risk we take when we believe that the creative process in science can be quantified aiming predictability.

Hirsch proposes an index to predict the future of scientists joining quality and quantity in a single number. As far as we are concerned this proposal of using mathematics probability or statistics aiming to predict future research achievement is not possible since the phase transition in a scientist career seems not to be predictable by an index or methodology as shown in this work through the example of the Nobel – and Ig Nobel – winner Andre Geim.

Although at first sight this could be negative, from a scientific point of view, since it indicates that one of the most important goals in Science is not achieved - the aim for stablishing models that can cover the phenomena, we can consider that it also preserves the element of surprise involved in the discovery, fundamental part and basis to Science.

As stated in (Braun 2012), having metrics as a single criterion to define research budget so as pay levels and pay rises is not connected to the real motivation of scientists and can lead to "bad science". On the other side, rewarding "linked to overall contributions" can really represent a great factor of motivations for scientists.

Nevertheless, why did *h* index widespread throughout the world and reach such a visibility? The first hypothesis is that this indicator fits the ideal of an epoch. That is, we believe that the *h* index appears at a time when the concept of a man who is always under evaluation is a reality as philosophical studies have pointed out (Zarka 2009).

The search for an ideal number portraits an attempt to find the best formula to the assessment of scientific activity. We try to find a quantitative value to classify a scientist considering him or her as an isolated person but we cannot forget that nowadays everybody is connected to a net being part of a community. Therefore, proposals like those seem to go beyond mathematics boundaries.

We tried to show in this work that the quantitative evaluation of research output is based on the development of a technological apparatus able to digitize scientific knowledge taking into account the citations found in papers.

Thus, focusing on Jorge Hirsch's proposal of *h* index we tried to show that mathematics when applied to a social field - in this case research activity - loses its objectivity, since any indicator results from human elaboration, therefore requiring a choice. This choice may be economical, political or even aesthetics. Therefore, indicators cannot bring with them the characteristics of universality and neutrality partially attributed to mathematical knowledge.

## Additional information

**Competing financial interests:** The authors declare no competing financial interests.